\documentclass[preprint,aps,prd,twocolumn,tightenlines,10pt,amsmath,amssymb]{revtex4}

\newcommand{\be}{\begin{eqnarray}}
\newcommand{\ee}{\end{eqnarray}}
\newcommand{\nn}{\nonumber~\\}

\begin{document}

\title{Saturation momentum at fixed and running QCD coupling}

\author{Dennis~D.~Dietrich}
\affiliation{Laboratoire de Physique Th\'eorique, Universit\'e Paris XI, Orsay, 
France}
\affiliation{The Niels Bohr Institute, Copenhagen, Denmark}


\begin{abstract}

A relationship, linking the saturation momentum in the case of fixed and
running QCD coupling, respectively, is derived from the Balitsky-Kovchegov
equation. It relies on the linear instability of the evolution equation in
the dilute regime. The leading orders of the saturation momenta are mapped
onto each other exactly. For subleading terms a qualitative correspondence
is achieved with a relative error going to zero for large rapidities. The
relationship can also be derived for the Balitsky-Kovchegov equation with a
cutoff accounting for low-density effects and is satisfied by the
corresponding isoline functions. Further implications arise for the
existence of travelling-wave solutions in the two situations. 

\end{abstract}

\maketitle


\section{Introduction and Results}

In deep-inelastic scattering the total cross-section $\sigma$ for the
scattering of a virtual photon with momentum $q$ on a proton with momentum
$p$ is a function of the virtuality $Q^2=-q^2$ and the rapidity $Y=\ln(1/x)$
with $x=Q^2/(2p\cdot q)$. At sufficiently small $x\lesssim 10^{-2}$ the
cross-section $\sigma$ becomes a function of the ratio of the virtuality
$Q^2$ and a function ${Q_s}^2(Y)$ of rapidity $Y$ called saturation momentum:
$\sigma=\sigma[Q^2/{Q_s}^2(Y)]$ \cite{dis}. This phenomenon is called
geometric scaling. Translated to the dipole scattering amplitude $N$, it
becomes a function of the difference between the logarithm $L$ of the square
of the momentum variable conjugate to the dipole size and the logarithm of
the saturation momentum: $N=N[L-\ln{Q_s}^2(Y)]$.

In quantum chromodynamics (QCD) the scattering of dipoles is described by
the Balitsky hierarchy \cite{b}. In the factorising limit it reduces to the
Balitsky-Kovchegov (BK) equation \cite{k}. In the low-density regime it in
turn simplifies to the Balitsky-Fadin-Kuraev-Lipatov (BFKL) equation
\cite{bfkl}.

In Ref.~\cite{mp} scaling solutions for the BK equation are identified at 
fixed coupling and with the second order ("diffusive") approximation 
\footnote{
The use of the approximate equation's solution is reasonable everywhere 
outside the 
deeply saturated regime. In the latter it goes to a constant while the 
solution of the exact equation continues to grow logarithmically \cite{k}. 
}
for the BFKL kernel. That investigation is based on knowledge about the
existence of travelling-wave solutions for the
Fisher-Kolmogorov-Petrovsky-Piskunov (FKPP) equation \cite{fkpp}.

This approach is not feasible for the running coupling BK equation because
it belongs to a different universality class. Different methods have to be
used, like a rescaling of the rapidity $Y$ \cite{iim}, introduction of
curved absorptive boundaries into the BFKL equation \cite{mt}, or use of a
travelling-wave ansatz \cite{mp2} 
\footnote{
All of these approaches work also in the fixed-coupling case.
}. 
Note that, in general, the "time" and "space" variables need not be linear
functions of the rapidity $Y$ and the momentum variable $L$.

Here a mapping between the saturation momentum at fixed and at running
coupling, respectively, is derived. The derivation is based on the
observation that in the scaling regime the saturation momentum characterises
an isoline of the amplitude $N$. It is derived by neglecting commutators of
the BFKL kernel and the running-coupling function acting on the amplitude
$N$. This procedure is justified by the fact that the BK equation is
linearly unstable with respect to small perturbations around $N=0$ (pulled
front). At this level of accuracy it is shown to hold for the saturation
momentum in both cases at leading order for large rapidities $Y$. For the
subleading terms the qualitative features are reproduced: They provide a
negative but small correction to the leading behaviour. The relative error
connected to the quantitative deviations vanishes for large rapidities $Y$.

As said instability is already present for the BFKL equation the
relationship also holds there. Actually the validity of the relation seems
to be widely independent of the detailed form of the equation of motion and
the form of the running coupling function's equivalent. It appears to hold
for a larger class of differential equations. In this context, the linear
instability seems to be a sufficient condition. The aforementioned quantitiative deviations are
non-universal in the sense that they depend on details of the equation of
motion which are independent of whether or not it is linearly unstable.

Furthermore, the relationship between the isolines for fixed and running
coupling can also be derived after a cutoff has been introduced into the
growth term of the BK equation in order to accomodate effects beyond the
mean-field approximation \cite{bd,ms,imm}. Even if, strictly speaking, the
thus modified BK equation is no longer linearly unstable against arbitrarily
small perturbations around $N=0$, the leading-order terms for large
rapidities $Y$ are again mapped onto each other exactly. The subleading
terms decay exponentially \cite{pvs}. For the approximation made in the
course of the relationship's derivation to hold, it suffices that the BK
equation with cutoff is unstable for sufficiently large perturbations.

The comparison of isoline plots \cite{iso}
obtained by solving the BK equation numerically \cite{num} for fixed and
running coupling respectively, according to the relations presented below,
would also provide interesting insights and cross checks.

In section \ref{relation} the mapping between the isolines for the BK
equation for fixed and running coupling, respectively, is derived. In
section \ref{comparison} the validity of the relationship is checked for the
leading (subsection \ref{leading}) and sub-leading terms (subsection
\ref{subleading}) of the saturation momentum. In subsection \ref{discussion} 
the reason for the accuracy of the relation is discussed. It is explained
how it can be generalised to a larger class of differential equations. 
Section \ref{lowdensity} treats the mapping for the BK equation with a
cutoff taking into account low-density effects. For convenience appendix 
\ref{FKPP} exposes details of the connection between the fixed coupling BK
equation and the FKPP equation.


\section{Relation\label{relation}}

The BK equation for the dipole forward-scattering amplitude as a function of 
the rapidity $Y$ and the momentum variable $L$ is given by: 
\be
\frac{\partial N}{\partial Y}
=
\bar{\alpha}
\left[
\chi\left(-\frac{\partial}{\partial L}\right)N
-
N^2
\right]
\label{bk}
\ee

\noindent
with the BFKL kernel:
\be
\chi(\gamma)=2\psi(1)-\psi(\gamma)-\psi(1-\gamma)
\label{bfklkernel}
\ee

\noindent
where $\psi(\gamma)$ is the digamma function---the logarithmic derivative
of the gamma function---and where the BFKL
eigenvalue-function with a differential operator in the argument is defined
via its series expansion around $\gamma_0\in]0;1[$.

The phenomenon of scaling manifests itself by the isolines of the amplitude
$N(Y,L)$ keeping their distance from each other in the direction of the 
variable $L$ constant if the rapidity $Y$ changes. In other words, the 
amplitude is only a function of the difference 
\mbox{$L-\ln Q_s(Y)^2$}. 
Consequently, in the scaling regime, the saturation momentum 
\mbox{$Q_s=Q_s(Y)$} characterises the isoline with 
$N[Y,Q_s(Y)]=N_s$ \footnote{The last condition defines this isoline also 
outside the scaling regime, although, strictly speaking, there, $Q_s$ does 
not deserve the name ``saturation momentum''}.
There, the equations describing two isolines differ merely by 
an additive constant.
 
Independent of the phenomenon of scaling, any isoline \mbox{$L_i=L_i(Y)$} of 
the amplitude $N(Y,L)$ satisfies the relation
\be
\frac{dL_i}{dY}
=
-
\frac{\partial N}{\partial Y}
\left(
\frac{\partial N}{\partial L_i}
\right)^{-1}.
\label{iso}
\ee

\noindent
For fixed QCD coupling, 
\mbox{$\partial N/\partial Y$} can been replaced by 
the right-hand side of (\ref{bk}). 
\be
\frac{dL_f}{dY}
=
-
\bar{\alpha}
\left[
\chi\left(-\frac{\partial}{\partial L_f}\right)N_f
-
{N_f}^2
\right]
\left(
\frac{\partial N_f}{\partial L_f}
\right)^{-1}.
\label{isof}
\ee

\noindent
The running coupling case is obtained through the replacement
\mbox{$
\bar{\alpha}\rightarrow(bL)^{-1}
$}:
\be
\frac{dL_r}{dY}
=
-
\frac{1}{bL_r}
\left[
\chi\left(-\frac{\partial}{\partial L_r}\right)N_r
-
{N_r}^2
\right]
\left(
\frac{\partial N_r}{\partial L_r}
\right)^{-1},
\label{isor}
\ee

\noindent
The constant $b$ is linked to the QCD-$\beta$-function and reads:
\mbox{$b=(11n_c-2n_f)/(12n_c)$}. $n_c$ stands for the number of colours and
$n_f$ for the number of (massless) flavours. 

Hence, division of Eq. (\ref{isof}) by $L_f$, taking into account the 
common initial condition
\be
N_f(Y_0,L)=N_0(L)=N_r(Y_0,L),
\ee

\noindent
yields: 
\be
\frac{1}{L_f}\frac{dL_f}{dY_0}
=
\frac{dL_r}{dY_0}.
\ee

\noindent
The constants $\bar{\alpha}$ and $b$ have been omitted for the sake of
clarity. It suffices to keep in mind to exchange the two in a comparison.

Analogous relations for the higher derivatives can be derived by
repetition of the above steps: After taking the derivatives of Eqs.
(\ref{isof}) and (\ref{isor}) with respect to the rapidity $Y$, replace the 
new occurrences of the derivative $\partial N/\partial Y$ on the right-hand side with the 
help of the BK equation (\ref{bk}). 
Subsequently, divide the expression obtained from Eq. (\ref{isof}) by 
$L_f$. This allows to identify the respective right-hand sides at $Y=Y_0$
up to terms 
involving commutators of the BFKL kernel with the running coupling function
acting on the amplitude $[\chi(-\partial_L),L^{-1}]_-N$. They are going to be
omitted in what follows. Why this is justified is going to be
investigated in subsection \ref{discussion}.

Together with the initial condition:
\be
L_f(Y_0)=L_0=L_r(Y_0)
\ee

\noindent
this hierarchy of equations can be summarised by:
\be
\left[
\frac{1}{L_f(Y_0)}\frac{d}{dY_0}
\right]^n 
L_f(Y_0)
=
\left[\frac{d}{dY_0}\right]^n L_r(Y_0)
\label{implicit}
\ee

\noindent
for all $n\in\mathbb{N}_0$. This defines all Taylor coefficients of 
$L_r(Y)$ at rapidity $Y=Y_0$ based on those of $L_f(Y)$. 

Inversely, by multiplying
with $L_r$ each time instead of dividing by $L_f$ one obtains all Taylor
coefficients of $L_r(Y)$ from those of $L_f(Y)$:
\be
\left[\frac{d}{dY_0}\right]^n L_f(Y_0)
=
\left[L_r(Y_0)\frac{d}{dY_0}\right]^n L_r(Y_0).
\label{implicitalternative}
\ee

Equation (\ref{implicit}) can be reexpressed as:
\be
L_r(Y_0+\delta Y)
=
\exp\left\{\delta Y\frac{1}{L_f(Y_0)}\frac{d}{dY_0}\right\}L_f(Y_0)
\label{explicit}
\ee

\noindent
and Eq. (\ref{implicitalternative}) accordingly as:
\be
L_f(Y_0+\delta Y)
=
\exp\left\{\delta Y L_r(Y_0)\frac{d}{dY_0}\right\}L_r(Y_0)
\label{explicitalternative}
\ee 

\noindent
for all $\delta Y$.
This can be verified by Taylor expansions around $\delta Y=0$. 
The exponentials in Eqs. (\ref{explicit}) and 
(\ref{explicitalternative}) are operators for conformal mappings. In the 
following calculations the interpretation as translation operators
is to be used. (Alternative computations could, for example, involve 
dilatation operators.) With the definitions:
\be
\frac{dZ_f}{dY}=L_f(Y),
\label{substitution}
\ee
  
\noindent
and
\be
\frac{dZ_r}{dY}=\frac{1}{L_r(Y)},
\label{substitutionalternative}
\ee

\noindent
respectively, these turn into:
\be
L_r(Y_0+\delta Y)
&=&
\exp\left\{\delta Y \frac{d}{dZ_f}\right\}L_f[Y_f(Z_f)]
=
\nn
&=&
L_f\{Y_f[Z_f(Y_0)+\delta Y]\}
\label{translation}
\ee

\noindent
and
\be
L_f(Y_0+\delta Y)
&=&
\exp\left\{\delta Y \frac{d}{dZ_r}\right\}L_r[Y_r(Z_r)]
=
\nn
&=&
L_r\{Y_r[Z_r(Y_0)+\delta Y]\}
\label{translationalternative}
\ee

\noindent
for all $\delta Y$.
$Y_f(Z_f)$ and $Y_r(Z_r)$ are the inverse functions of
$Z_f(Y)$ and $Z_r(Y)$, respectively, as defined in Eqs.
(\ref{substitution}) and (\ref{substitutionalternative}).
Eqs. (\ref{translation}) and (\ref{translationalternative}) provide 
a direct link between the isolines of the solutions for the equations of 
motion in the fixed 
and the running coupling case. They hold irrespective of whether the exact or
an approximative expression, for example, the second-order expansion 
around $\gamma=\gamma_c$, is used for the BFKL 
kernel (\ref{bfklkernel}).


\section{Comparison\label{comparison}}

For large rapidities $Y$, the universal expression for the (logarithm of
the) saturation momentum---up to an arbitrary additive constant---in the
case of fixed coupling available in the literature reads \cite{mt,mp2,mp3}:
\be
L_f(Y)
=
\bar{\alpha}\frac{\chi(\gamma_c)}{\gamma_c}Y
-
\frac{3}{2\gamma_c}\ln Y
-
\frac{3}{{\gamma_c}^2}
\sqrt{\frac{2\pi}{\bar{\alpha}\chi^{\prime\prime}(\gamma_c)}}
\sqrt{\frac{1}{Y}}.
\label{lf}
\ee

\noindent
The BFKL kernel (\ref{bfklkernel}) has been expanded around
\mbox{$\gamma_0=\gamma_c$} which solves the equation
\mbox{$\gamma_c\chi^\prime(\gamma_c)=\chi(\gamma_c)$} and has the numerical
value \mbox{$\gamma_c=0.6275...$} \cite{mp2}.
The third term is only given in \cite{mp3} and there the error is of the order
\mbox{${\cal O}(Y^{-1})$}.
For the running coupling case the two leading terms are known \cite{mt,mp2}:
\be
L_r(Y)
=
\sqrt{\frac{2\chi(\gamma_c)}{b\gamma_c}Y}
+
\frac{3}{4}
\left[
\frac{\chi^{\prime\prime}(\gamma_c)}{\sqrt{2b\gamma_c\chi(\gamma_c)}}
\right]^{\frac{1}{3}}
\xi_1 Y^{\frac{1}{6}}
\label{lr}
\ee

\noindent
with $\xi_1=-2.338...$ being the rightmost zero of the Airy function 
${\rm Ai}$. The universally valid expression, i.e., independent of the 
initial conditions, can at most include terms that decrease more slowly than
\mbox{$Y^{-1/2}$}. Like in the case of fixed QCD coupling, this can be 
checked through a shift $\Delta Y$ of the
rapidity $Y$ in the above equation and subsequent expansion around 
\mbox{$\Delta Y=0$} \cite{mp3}. Thus, in general, the rapidity variable $Y$ 
in the last two equations differs by this additive constant $\Delta Y$.


\subsection{Leading order\label{leading}}

Starting out with the first term in Eq. (\ref{lf}) one finds from Eq.
(\ref{substitution}):
\be
Z_f(Y)
=
\bar{\alpha}\frac{\chi(\gamma_c)}{2\gamma_c}{Y}^2.
\label{zfone}
\ee

\noindent
Inversion yields:
\be
Y_f(Z_f)
=
\sqrt{\frac{2\gamma_c}{\bar{\alpha}\chi(\gamma_c)}Z_f}
\label{inverseleading}
\ee

\noindent
Evaluation at \mbox{$Z_f(Y_0)+\delta Y$} leads to:
\be
Y_f[Z_f(Y_0)+\delta Y]
=
\sqrt{\frac{2\gamma_c}{\bar{\alpha}\chi(\gamma_c)}[Z_f(Y_0)+\delta Y]}.
\label{evaluateleading}
\ee

\noindent
Replacing the rapidity 
$Y$ in the first term of Eq. (\ref{lf}) by the right-hand side of the 
previous expression and afterwards $\bar{\alpha}$ by $b^{-1}$ yields:
\be
\sqrt{\frac{2\chi(\gamma_c)}{b\gamma_c}(Y_0+\delta Y+\Delta Y)}
=
L_r(Y_0+\delta Y),
\label{replace}
\ee

\noindent
where constants have been absorbed in $\Delta Y$. Thus, to
this order, Eq. (\ref{translation}) is satisfied. With analogous
calculations also Eq. (\ref{translationalternative}) can be verified to
leading order.


\subsection{Subleading terms\label{subleading}}

With the first two terms in Eq. (\ref{lf}) one finds from Eq.
(\ref{substitution}):
\be
Z_f(Y)
=
\bar{\alpha}\frac{\chi(\gamma_c)}{2\gamma_c}{Y}^2
+
\frac{3}{2\gamma_c}(1-\ln Y)Y,
\label{zfsubleading}
\ee

\noindent
the exact inverse of which cannot be given analytically.
The factor \mbox{$(1-\ln Y)$} varies slowly as compared to powers of $Y$.
Regarding it as fixed in every step, one can determine the inverse 
iteratively. Starting
out with \mbox{$\ln Y_f^{(0)}(Z_f)=1$} leads to Eq. (\ref{inverseleading}) 
for \mbox{$Y_f^{(1)}(Z_f)$}. Replacing $\ln Y$ in the logarithm in Eq.
(\ref{zfsubleading}) by the preliminary result \mbox{$\ln Y_f^{(1)}(Z_f)$} [in
general \mbox{$\ln Y_f^{(n)}(Z_f)$}], 
subsequent inversion, and selection of the positive solution in every step 
finally leads to the recursion relation:
\be
Y_f^{(n+1)}(Z_f)
=
-\frac
{3\left[1-\ln Y_f^{(n)}(Z_f)\right]}
{2\bar{\alpha}\chi(\gamma_c)}
+
\nn
+
\sqrt{
\left\{
\frac
{3\left[1-\ln Y_f^{(n)}(Z_f)\right]}
{2\bar{\alpha}\chi(\gamma_c)}
\right\}^2
+
[Y_f^{(1)}(Z_f)]^2
}
\nn
\label{recursion}
\ee

\noindent
It shows that the dominant behaviour for large $Z_f$---and hence large
$Y_f^{(1)}(Z_f)$---can be obtained after a finite number of iterations.
The dominant terms are given by:
\be
Y_f^{(n+1)}(Z_f)
=
Y_f^{(1)}(Z_f)
+
\frac{3}{2\bar{\alpha}\chi(\gamma_c)}\ln Y_f^{(n)}(Z_f)
-
\nn
-
\frac{1}{2}
\left[
\frac{3}{2\bar{\alpha}\chi(\gamma_c)}
\right]^2
\frac
{[\ln Y_f^{(n)}(Z_f)]^2}
{Y_f^{(1)}(Z_f)}
+
{\cal O}\left[\frac{1}{Y_f^{(1)}(Z_f)}\right],
\ee

\noindent
where constant terms have been omitted. Replacing the rapidity $Y$ in the 
first two terms of Eq. (\ref{lf}) by the right-hand side of the previous 
expression yields:
\be
\bar{L}_r
&=&
\bar{\alpha}\frac{\chi(\gamma_c)}{\gamma_c}
Y_f^{(1)}(Z_f)
-
\frac{9}{8}
\frac{1}{\bar{\alpha}\chi(\gamma_c)\gamma_c}
\frac
{[\ln Y_f^{(1)}(Z_f)]^2}
{Y_f^{(1)}(Z_f)}
+
\nn
&&+
{\cal O}
\left[
\frac{\ln Y_f^{(1)}(Z_f)}{Y_f^{(1)}(Z_f)}
\right],
\label{lrbar}
\ee

\noindent
which can already be obtained from the second-order result, i.e., from
Eq. (\ref{recursion}) with $n=1$.
Evaluation at \mbox{$Z_f(Y_0)+\delta Y$} leads to the replacement:
\be
&&
Y_f^{(1)}(Z_f)
\rightarrow
Y_f^{(1)}[Z_f(Y_0)+\delta Y]
=
\nn
&=&
\frac{\gamma_c}{\bar{\alpha}\chi(\gamma_c)}
\sqrt{\frac{2\chi(\gamma_c)}{b\gamma_c}(Y_0+\delta Y+\Delta Y)},
\label{replacement}
\ee

\noindent
which coincides with Eq. (\ref{evaluateleading}). Finally,
$\bar{\alpha}$ has to be replaced by $b^{-1}$. 

Again, the leading term of Eq. (\ref{lr}) is reproduced. The first
subleading terms in Eqs. (\ref{lr}) and (\ref{lrbar}) together with 
(\ref{replacement}) do not coincide exactly. However, they are similar 
qualitatively. Through their inclusion with the leading term, $L_r$ and 
$\bar{L}_r$ are both diminished. The relative error 
vanishes for large rapidities $Y$ like:
$
(\bar{L}_r-L_r)/(\bar{L}_r+L_r)
\sim
Y^{-1/3}.
$

In principle, it is possible to base the above comparison on Eq.
(\ref{translationalternative}). While the integral required for solving
Eq. (\ref{substitutionalternative}) is still known analytically for
$L_r(Y)$ given by Eq. (\ref{lr}), the inversion of the resulting
expression is more cumbersome than for Eq. (\ref{zfsubleading}).


\subsection{Discussion\label{discussion}}

As demonstrated in the previous subsection, neglecting the commutator
$[\chi(-\partial_L),L^{-1}]_-N$ leads to a mapping exact at leading order
and with subleading deviations whose relative error vanished for large
rapidities. In what follows it shall be discussed why this is the case.

Omitting said commutator is equivalent to approximating the prefactors of
the $m^{th}$ derivatives with respect to the momentum variable $L$ occuring
on the running-coupling side during the relationship's derivation by the
term dominant for large $L$:
$[L^{-1}+{\cal O}(L^{-2})]{\partial_L}^mN\approx L^{-1}{\partial_L}^mN$. 
Subsequently one would have to justify why $L$ is effectively so large that
the above steps are feasible. One is tempted to bring forward the fact that
saturation physics is protected from the influence of the infrared, i.e.,
from small $L$. However, through the repetition of the identical steps the
above relationship can also be derived for the BFKL equation and in the BFKL
equation no saturation effects are encoded.

As this line of arguments is not conclusive let us investigate how
shifting the BFKL kernel by an additive constant
$\bar{\chi}(\gamma)=\chi(\gamma)+\delta$ does affect the expression for the
saturation momentum. First for arbitrary values of the shift
$\delta\in\mathbb{R}$ the modified critical value $\bar{\gamma}_c$ for the
argument $\gamma$, defined through
$\bar{\chi}(\bar{\gamma}_c)=\bar{\gamma}_c\bar{\chi}^\prime(\bar{\gamma}_c)$
obeys $0<\bar{\gamma}_c<1$. Therefore the solution stays always in the
supercritical regime of the FKPP equation ${\bar{\gamma}_c}^{-1}>1$
\footnote{In order to clarify the necessary connections 
Appendix \ref{FKPP} displays the mapping that leads to the FKPP equation.}, 
i.e., it has always a universal travelling-wave solution 
\cite{mp,mp2}. Hence one can explore the effect of the shift $\delta$ 
directly with the help of the expression for the saturation momentum in Eq. 
(\ref{lf}). 

For $\delta>-4\ln 2$, $\bar{\chi}(\gamma)$ remains 
positive definite and the same qualitative asymptotic behaviour is
retained, although reaching the asymptotic regime might require extremely
large rapidities $Y$ if $\bar{\chi}(\bar{\gamma}_c)\ll 1$.

The picture changes for $\delta =-4\ln 2$, where
$\bar{\gamma}_c=\frac{1}{2}$ and $\bar{\chi}(\bar{\gamma}_c)=0$, i.e, the
minimum of the kernel becomes zero. The term proportional to the rapidity
$Y$ is absent. As explained above, by shifting the BFKL kernel one stays in 
the supercritical regime of the FKPP equation. However, the $L$ and the $x$
axes are not parallel to each other. In this particular situation $x\sim 2t$
is mapped exactly onto $L=\mathrm{const.}$ (see Appendix \ref{FKPP}), whence
the linear term of the saturation momentum vanishes although the FKPP
equation has a supercritical travelling-wave solution. In other words, for
$\delta =-4\ln 2$ the BK equation is not linearly unstable for small
perturbations around $N=0$ even if the FKPP equation is.

The effect on the expression for running coupling (\ref{lr}) is more
drastic. While the term proportional to the square root of the rapidity $Y$
vanished in unison with the linear term in the fixed coupling case, the
prefactor of the second addend diverges like 
$\sim{\bar{\chi}(\gamma_c)}^{-1/6}$.
Hence, in the running-coupling case the previous description breaks down
altogether.

Proceeding to $\delta <-4\ln 2$, leads to $\bar{\chi}(\gamma)$ also taking
negative values and especially to a negative critical value
$\bar{\chi}(\bar{\gamma}_c)<0$.  Thus the term in the fixed-coupling saturation
momentum proportional to the rapidity $Y$ reappears but with a negative
sign. This means that for growing rapidities $Y$ the amplitude decreases.
With this kernel the BK equation is stable against perturbations around
$N=0$. In this range, according to Eq. (\ref{lr}), the saturation momentum
in the running-coupling case would even be complex. This shows that the
latter case would have to be investigated anew.

Summarising, the term of the differential equation important for the
instability around $N=0$ and hence for the mapping to work is the one
originating from the critical value of the kernel 
${\chi}({\gamma}_c)$.  Preserving
only this term in the BK equation and obtaining the relationship between the
isolines for the different couplings leads exactly to the previous results,
because in this limit the crucial commutator vanishes
$[\chi(\gamma_c),L^{-1}]N=0$. The more general derivation in section
\ref{relation} resums additional terms which lead to the same overall
behaviour.

As an outlook, based on the discussions in this subsection the present
approach should work for any pair of differential equations which are
linearly unstable around the dilute state, widely independent of the details
of the remaining terms and the detailed structure of the equivalent of the
running coupling-constant.  The main step for adapting to another
situation is replacing the right-hand sides of Eqs.
(\ref{substitutionalternative}) or (\ref{substitution}), respectively, by
the new running coupling function $\bar{\alpha}(L)$ or its reciprocal
$[\bar{\alpha}(L)]^{-1}$, respectively.


\section{Low-density effects\label{lowdensity}}

The BK equation describes the mean-field limit of the Balitsky
hierarchy. The mean-field approximation is best satisfied in the dense
regime and least in the dilute. There fluctuations are important. As
mentioned above, for the relevant boundary conditions, the BK equation
describes the propagation into a linearly unstable state. This leads to a
high sensitivity of the solution to modifications at the toe of the front.
In Refs.~\cite{bd,imm} it has been demonstrated that the principal correction 
to the front propagation speed can be simulated in a deterministic manner by
cutting off the growth term for small values of the amplitude. For example
in the diffusive approximation to the BK equation this leads to the
following modified equation of motion:
\be
\frac{\partial N}{\partial Y}
=
\bar{\alpha}
\left[
\chi_2\frac{\partial^2 N}{\partial L^2}
+
\chi_1\frac{\partial N}{\partial L}
+
\left(
\chi_0N
-
N^2
\right)
c(N)
\right]
\label{bkcut}
\ee
with the replacement $\bar{\alpha}\rightarrow(bL)^{-1}$ for the case of
running coupling and where the coefficients $\chi_i$, $i\in\{1,2,3\}$ are 
given in Appendix \ref{FKPP} and with the cutoff function:
\be
c(N)=\theta\left(N-{\alpha_S}^2\right).
\ee
Note that $1/{\alpha_S}^2$ equals the number of dipoles at saturation
wherefore ${\alpha_S}^2$ gives the corresponding step height.

Carrying out the steps of the derivation beginning with Eq. (\ref{iso}) but
this time for the modified BK equation (\ref{bkcut}) instead of its standard
form (\ref{bk}) yields again Eqs. (\ref{translation}) and
(\ref{translationalternative}). Looking at the modified expressions for the
saturation momentum one sees that Eqs. (\ref{translation}) and
(\ref{translationalternative}) are satisfied exactly: At fixed coupling the
leading term of the isoline equation reads \cite{bd,ms,imm}:
\be
L_f(Y)
=
\bar{\alpha}
\left[
\frac{\chi(\gamma_c)}{\gamma_c}
-
\frac{\pi^2}{2}
\frac{\gamma_c\chi^{\prime\prime}(\gamma_c)}{\ln^2(1/{\alpha_s}^2)}
\right]
Y,
\ee
with exponentially small subleading terms \cite{pvs}. At running coupling
one finds \cite{ms}:

\be
L_r(Y)
=
\sqrt{
\frac{2}{b}
\left[
\frac{\chi(\gamma_c)}{\gamma_c}
-
\frac{\pi^2}{2}
\frac{\gamma_c\chi^{\prime\prime}(\gamma_c)}{\ln^2(1/{\alpha_s}^2)}
\right]
Y
}.
\ee

The introduction of the cutoff removes the linear instability of the BK 
equation with respect to perturbations smaller than the threshold
${\alpha_S}^2$. However, the instability for perturbations larger than
${\alpha_s}^2$ is sufficient to ensure that the modified expressions for the
saturation momentum are qualitatively similar to those of the unmodified
version. Consequently the mapping still works.

In this last context the term "saturation momentum" is avoided on purpose
because after the inclusion of stochastic effects the solutions of the BK
equation does no longer provide the observable amplitude but the amplitude
for the scattering on a given partonic realisation of the target
\cite{ms,imm}. The physical amplitude is obtained by means of an ensemble
average \cite{imm} accounting for the non-vanishing variance of the front
position \cite{bd}. Then the physical amplitude does no longer show geometric
scaling \cite{ms,imm}.


\section*{Acknowledgments}

The author feels particularly indebted to Gregory Kor\-chem\-sky for giving
helpful and informative answers to his questions. The author would also like
to thank Habib Aissaoui, Yacine Mehtar-Tani, Joachim Reinhardt, and
Dominique Schiff for lively discussions. This work has been supported
financially by the German Academic Exchange Service (DAAD).


\appendix

\section{Mapping: FKPP$\leftrightarrow$BK\label{FKPP}}

The fixed-coupling BK equation for $N=N(Y,L)$ in the second-order 
("diffusive") approximation:
\be
\frac{\partial N}{\partial Y}
=
\bar{\alpha}
\left[
\chi_2\frac{\partial^2L}{\partial L^2}
+
\chi_1\frac{\partial N}{\partial L}
+
\chi_0N
-
N^2
\right],
\ee
with:
\be
\chi_0&=&{\gamma_c}^2\chi^{\prime\prime}(\gamma_c)/2,
\nn
\chi_1&=&\gamma_c\chi^{\prime\prime}(\gamma_c)+\chi(\gamma_c)/\gamma_c,
\nn
\chi_0&=&\chi^{\prime\prime}(\gamma_c)/2,
\ee
is mapped onto the FKPP equation for $u=u(t,x)$:
\be
\frac{\partial u}{\partial t}
=
\frac{\partial^2u}{\partial x^2}
+
u
-
u^2,
\ee
by the relation:
\be
N(Y,L)=[{\gamma_c}^2\chi^{\prime\prime}(\gamma_c)/2]\times u[t(Y),x(Y,L)],
\ee
with:
\be
t
&=&
\bar{\alpha}Y\times[{\gamma_c}^2\chi^{\prime\prime}(\gamma_c)/2],
\nn
x
&=&
\bar{\alpha}Y\times[{\gamma_c}^2\chi^{\prime\prime}(\gamma_c)+\chi(\gamma_c)]
+
L\times\gamma_c.
\ee



\end{document}